\documentclass[%
 preprint,
superscriptaddress,
longbibliography,
amsmath,amssymb,
aps,
prb,
nofootinbib,
]{revtex4-1}

\usepackage{graphicx}
\usepackage{dcolumn}
\usepackage{bm}
\usepackage{hyperref}

\usepackage[allowzeroexp=true,load={abbr,addn}]{siunitx}
\usepackage{amsmath}
\usepackage{natbib}
\usepackage[english]{babel}

\begin{document}


\title{Towards Time-Resolved Atomic Structure Determination by X-Ray Standing Waves at a Free-Electron Laser}

\author{Giuseppe Mercurio}
\email{giuseppe.mercurio@xfel.eu}
\affiliation{Department Physik, Universit\"{a}t Hamburg, Luruper Chaussee 149, D-22761 Hamburg, Germany}
\affiliation{Center for Free-Electron Laser Science, Luruper Chaussee 149, D-22761 Hamburg, Germany}
\author{Igor A. Makhotkin}
\author{Igor Milov}
\affiliation{Industrial Focus Group XUV Optics, MESA+ Institute for Nanotechnology, University of Twente, Drienerlolaan 5, 7522 NB Enschede, Netherlands}
\author{Young Yong Kim}
\affiliation{Deutsches Electronen-Synchrotron DESY, Notkestrasse 85, D-22607 Hamburg, Germany}
\author{Ivan A. Zaluzhnyy}
\affiliation{Deutsches Electronen-Synchrotron DESY, Notkestrasse 85, D-22607 Hamburg, Germany}
\affiliation{National Research Nuclear University MEPhI (Moscow Engineering Physics Institute), Kashirskoe shosse 31, 115409 Moscow, Russia}
\author{Siarhei Dziarzhytski}
\affiliation{Deutsches Electronen-Synchrotron DESY, Notkestrasse 85, D-22607 Hamburg, Germany}
\author{Lukas Wenthaus}
\affiliation{Department Physik, Universit\"{a}t Hamburg, Luruper Chaussee 149, D-22761 Hamburg, Germany}
\affiliation{Center for Free-Electron Laser Science, Luruper Chaussee 149, D-22761 Hamburg, Germany}
\author{Ivan A. Vartanyants}
\affiliation{Deutsches Electronen-Synchrotron DESY, Notkestrasse 85, D-22607 Hamburg, Germany}
\affiliation{National Research Nuclear University MEPhI (Moscow Engineering Physics Institute), Kashirskoe shosse 31, 115409 Moscow, Russia}
\author{Wilfried Wurth}
\affiliation{Department Physik, Universit\"{a}t Hamburg, Luruper Chaussee 149, D-22761 Hamburg, Germany}
\affiliation{Center for Free-Electron Laser Science, Luruper Chaussee 149, D-22761 Hamburg, Germany}
\affiliation{Deutsches Electronen-Synchrotron DESY, Notkestrasse 85, D-22607 Hamburg, Germany}

\date{\today}

\begin{abstract}
We demonstrate the structural sensitivity and accuracy of the standing wave technique at a high repetition rate free-electron laser, FLASH at DESY in Hamburg, by measuring the photoelectron yield from the surface SiO$_2$ of Mo/Si multilayers. These experiments open up the possibility to obtain unprecedented structural information of adsorbate and surface atoms with picometer spatial accuracy and femtosecond temporal resolution. This technique will substantially contribute to a fundamental understanding of chemical reactions at catalytic surfaces and the structural dynamics of superconductors.
\begin{description}

\item[PACS numbers]
68.49.Uv, 41.60.Cr, 68.65.Ac

\end{description}
\end{abstract}

\pacs{}
\maketitle



\section{Introduction} \label{Introduction}

The use of renewable energies for heterogeneous catalysis imposes the understanding of catalytic processes under dynamic reaction conditions. To achieve this goal there is a need of time-resolved spectroscopy measurements, predictive theory and the development of new catalysts \cite{kalz_future_2017}. With the advent of x-ray free-electron lasers (XFEL) \cite{ackermann_operation_2007,emma_first_2010,ishikawa_compact_2012, allaria_highly_2012,Altarelli2006}, delivering femtosecond, extremely brilliant, and coherent pulses in the soft and hard x-ray range, it became possible to explore the ultrafast dynamics of heterogeneous catalysis using a pump-probe approach \cite{ostrom_probing_2015, larue_real-time_2017}. Optical laser pump pulses are absorbed at the catalyst surface and trigger the reaction by electronic or phononic excitations \cite{frischkorn_femtochemistry_2006}. XFEL probe pulses are used to measure time-resolved x-ray absorption and emission spectra. In this way several elementary processes, essential for understanding more complex chemical reactions, were unveiled: breaking of the bond between CO molecules and a Ru surface \cite{dellangela_real-time_2013}, transient excitation of O atoms out of their ground adsorption state \cite{beye_chemical_2016}, transient states of CO oxidation \cite{ostrom_probing_2015}, and hydrogenation reactions \cite{larue_real-time_2017}. The interpretation of these spectroscopic data relies on density functional theory (DFT) calculations. Only the comparison of measured and simulated data allows to sketch the time evolution of a chemical reaction \cite{ostrom_probing_2015}, as depicted in Fig.\ref{Figure-1}(a). At the same time, a direct structural information on the position of atoms and molecules during the reaction is still missing.


Time-resolved structures of sample surfaces can be obtained in principle by means of low energy electron diffraction (LEED) \cite{vogelgesang_phase_2017}, reflection high energy electron diffraction (RHEED) \cite{zewail_4d_2006,frigge_optically_2017} or surface x-ray diffraction \cite{gustafson_high-energy_2014}. However, all these methods require lateral long range order of the structure to be resolved. In the case of atoms and molecules involved in chemical reactions at surfaces this requirement may not be fulfilled \cite{mieher_bimolecular_1993}. Therefore, to measure the time-resolved structure of reactants and catalysts as the reaction proceeds at the surface, we propose to combine photoelectron spectroscopy with the structural accuracy of the x-ray standing wave (XSW) technique \cite{zegenhagen_XSW_Book_2013, zegenhagen_surface_1993, vartanyants_theory_2001, woodruff_surface_2005} and the time resolution provided by an XFEL. In this way we can obtain at the same time sensitivity to the chemical environment of the reactants, by photoelectron spectroscopy (e.g. Refs.\citep{FuggleSS75,lizzit_surface_2001,bjorneholm_overlayer_1994}), and to their position along the Bragg diffraction vector $\mathbf{H}$, by the standing wave technique (Fig. \ref{Figure-1}(a)). In fact, XSW proved already to be an ideal tool to determine position and geometry of adsorbates at metal surfaces \cite{woodruff_surface_2005,mercurio_adsorption_2013,mercurio_x-ray_2014,gerlach_orientational_2011}. Importantly, the predictive quality of DFT calculations will enormously profit from the experimental structural benchmark provided by time-resolved XSW data, leading ultimately to a better understanding of the fundamental processes in heterogeneous catalysis. 


In this pioneering experiment, we demonstrate the structural sensitivity and accuracy of the XSW technique combined with the ultrashort pulses of an XFEL. The XSW forms in the region of spatial overlap between two coherently coupled incoming and Bragg-diffracted x-ray plane waves \cite{zegenhagen_XSW_Book_2013}. This results in a periodic modulation of the x-ray intensity (with period $d_{SW}$, Fig. \ref{Figure-1}(b)) along the z direction (perpendicular to the reflecting planes) described by the following equation \cite{zegenhagen_surface_1993}:
\begin{equation} \label{eq_1}
I_{SW}\left( z, \theta \right) = 1 + R\left(\theta\right) + 2 \sqrt{R\left(\theta\right)} \cos\left( \phi\left(\theta\right) - 2\pi\frac{z}{d_{SW}} \right),
\end{equation}
where $R\left(\theta\right)$ is the sample reflectivity and $\phi\left(\theta\right)$ the phase of the ratio $E_\mathrm{H} / E_0 \left(\theta\right) = \sqrt{R\left(\theta\right)} \exp\left( i\phi\left(\theta\right) \right)$, with $E_0$ and $E_\mathrm{H}$ the complex electric field amplitude of the incoming and Bragg-diffracted electromagnetic waves. Note that both $R\left(\theta\right)$ and $\phi\left(\theta\right)$ are functions of the normal angle of incidence $\theta$ (Figs. \ref{Figure-2}(c) and \ref{Figure-5}).


The main interest of this technique lies in the inelastic scattering of the x-ray standing wave from atoms that work as a probe, leading to photoelectron or x-ray fluorescence yield. The strength of this scattering signal is proportional to the intensity of the XSW at the position of the emitting atoms. Thus, by moving the standing wave in space it is possible to obtain information about the location of the emitters along the perpendicular direction to the Bragg planes, with a spatial accuracy of about $0.01$ $d_{SW}$. In fact, in an XSW experiment as the normal angle of incidence $\theta$ of the incoming x-ray wave varies through the Bragg condition, the phase $\phi$ changes by $\pi$, thus the standing wave shifts along the $\mathrm{z}$ direction by $d_{SW}/2$. Typically the position of light atoms, predominantly present in the reactants, is monitored by the photoelectron signal due to the larger cross-section as compared to fluorescence \cite{zegenhagen_surface_1993}. Therefore, performing XSW experiments combined with photoelectron spectroscopy at an XFEL on chemical reactions at single crystal catalysts with $d_{SW}$ of about few \si{\angstrom} (using photon energy of few keV) may deliver structural information of the reactants with an unprecedented high spatial accuracy much below $1$ \si{}{\angstrom} \cite{mercurio_adsorption_2013} and femtosecond temporal resolution.


\section{Experimental details} \label{Experimental details}

Typically XSW experiments are carried out at synchrotron radiation facilities in order to profit from the photon energy tunability in the soft and hard x-ray range, allowing to match the period of multilayers \cite{libera_x-ray_2004, libera_direct_2005, Bedzyk13Book, yang_probing_2008, fadley_hard_2013} and single crystals \cite{woodruff_surface_2005}, and from the high flux in a small bandwidth that enables a fine scan of the Bragg condition. At the same time, the narrow bandwidth $\Delta \lambda$ ensures that the longitudinal coherence $l_c \propto \lambda^2 / \Delta \lambda$ \cite{Goodman85StatisticalOptics} is much larger than the optical path length difference between the two interfering waves. All these advantages are preserved at free-electron laser facilities. In addition, femtosecond FEL pulses enable studies of ultrafast dynamics up to few tens of femtosecond which could not be reached by $\sim 100$ \si{\pico\second} synchrotron pulses. At the same time, when measuring photoelectron spectra at an XFEL, due to the high intensity and ultrashort x-ray pulses, vacuum space-charge effects need to be considered \cite{pietzsch_towards_2008,hellmann_time-resolved_2012}. To avoid them, the XFEL intensity needs to be reduced, while preserving the short pulse duration, leading to a detection of about one electron per XFEL pulse (limited by our spectrometer, see section \ref{Experimental chamber}). As a consequence, in order to measure time-resolved photoelectron spectra to probe sub-\si{\pico\second} to \si{\pico\second} dynamics with good statistics and in a reasonable amount of time, a high repetition rate XFEL is necessary.


\subsection{Free-electron laser parameters} \label{FLASH parameters}

The XSW experiment was performed at the Free-Electron LASer FLASH at DESY in Hamburg \cite{ackermann_operation_2007,tiedtke_soft_2009}, the only high repetition rate XFEL in operation at the time of the experiment. FLASH was operated in the multibunch mode, delivering pulse trains with a repetition rate of $10$ \si{\Hz}. Each bunch train consisted of 400 [320] pulses with photon energy of $93$ \si{\eV} (or $91.7$ \si{\eV}, Table \ref{table-01-Filter} in Appendix \ref{electron-yield-normalization}) and an intrabunchtrain repetition rate of $1$ \si{\MHz} for FLASH1 operation only [parallel operation of FLASH1 and FLASH2 sharing the same LINAC \cite{FaatzNJoP16}]. 

When measuring photoelectron spectra, space-charge effects need to be taken into account. In fact, if too many photoelectrons are emitted in a small area and in a short femtosecond time, the resulting photoemission spectrum will be energy shifted and broadened due to Coulomb repulsion \cite{pietzsch_towards_2008,hellmann_time-resolved_2012}. To avoid this, the intensity of the FEL pulses was reduced by a gas attenuator filled with \SI{2.7e-2}{\milli\bar} Xe gas as well as by solid filters (see Appendix \ref{electron-yield-normalization}). The monochromator was set to the first order diffraction from the 200 lines/mm plane grating with a fix-focus constant ($c_{ff}$) of $1.25$ and the exit slit was set to $100$ \si{\um} to have $40$ \si{\milli\electronvolt} energy bandwidth. As a result, the number of photons per pulse at the sample was on average \SI{5e+7} with a beam size ($\approx 30$ \si{\cm} behind the XFEL focus) of $150 - 200$ \si{\um} FWHM, therefore the corresponding fluence was $< 1$ \si{\micro\joule\per\square\cm}. This fluence was 5 orders of magnitude smaller than the single shot damage threshold of Mo/Si ML at normal incidence ($83$ \si{\milli\joule\per\square\cm}) \cite{MakhotkinJoSR18}. Therefore radiation induced sample damage could be excluded. Moreover, initial XFEL pulses $\approx 60$ \si{\fs} FWHM long were elongated to $\approx 200$ \si{\fs} FWHM due to the pulse front tilt at the monochromator grating.

\subsection{Experimental chamber} \label{Experimental chamber}

The measurements were carried out in the experimental chamber WESPE (Wide-angle Electron SPEctrometer) equipped with a vertical manipulator to tune the angle of incidence $\theta$, and the electron time-of-flight spectrometer THEMIS 1000 (SPECS), provided with a four-quadrant delay line detector (Surface Concepts), to measure photoelectron spectra. The spectrometer was set to measure electrons of kinetic energy 63 \si{eV}, pass energy $40$ \si{eV} and with an acceptance angle of $\pm 6\degree$. Given these settings, and the FEL attenuation needed to avoid space charge effects, on average $0.5 - 1.0$ electron per FEL pulse was detected.


\subsection{Samples} \label{Samples}

Since FLASH operates in the soft x-ray range ($\lambda = 4.2 - 52$ \si{\nm}), the periodic structure generating the standing wave had to have a period of comparable size. Our samples were Mo/Si multilayers (ML), consisting of 50 Mo/Si bilayers deposited on a super-polished Si substrate by means of sequential magnetron sputtering of Mo and Si in Ar atmosphere \cite{LouisPSS11}. The Si substrates were placed on a rotating substrate holder above the magnetron, such that all installed substrates could be coated at the same time and all coated layers were identical. The thickness of each layer was controlled by pre-calibrated sputtering time leading to a nominal multilayer period of $d_{ML} = 7.3$ \si{\nm} (Fig. \ref{Figure-1}(c)). To match the first order Bragg condition $2 d_{ML} \sin \left( \pi/2 - \theta \right) = \lambda$, FLASH was tuned to the wavelength $\lambda = 13.5 $ \si{\nm} and the normal angle of incidence of the maximum reflectivity was $\theta_{max} = 17.5$\si{\degree} (Fig. \ref{Figure-5}). In order to demonstrate the structural sensitivity of the XSW technique using an XFEL, we employed 4 ML samples terminated with the top Si layer of different nominal thickness $d_{\mathrm{Si}}^{top}$. After the deposition of the last Mo layer a system of masks was used to enable coating of the top Si layer with different thicknesses $d_\mathrm{Si}^{top}$. As a result we obtained four identical periodic Mo/Si MLs terminated with nominal top Si layers of thickness $2.0$ \si{\nm}, $2.8$ \si{\nm}, $3.6$ \si{\nm}, and $4.3$ \si{\nm}, referred to as sample 1, 2, 3, and 4 respectively.

As the Si-terminated ML samples were exposed to air, a native SiO$_2$ layer of $d_{\mathrm{SiO_2}} = 1.2$ \si{\nm} formed at the surface (see Appendix \ref{SiO2 thickness}). This led to 4 different distances of the surface oxide from the underlying identical periodic structure. In our XSW experiments we measured the photoelectron yield of O2s core level originating from the O atoms located at the surface of the SiO$_2$ layer (see Fig. \ref{Figure-1}(c)) as a function of the incident angle $\theta$. In this way we probed the position of the surface relative to the standing wave modulation and demonstrated the structural sensitivity of the XSW technique at an XFEL source. Based on this, it will be possible to measure changes in the electronic structure of atoms with picometer spatial accuracy at femtosecond time resolution.


\section{Results and discussion} \label{Results and discussion}


\subsection{Photoelectron spectra} \label{Photoelectron spectra}

A typical photoelectron spectrum measured on one of our ML samples is shown in Fig. \ref{Figure-2}(b). The most intense peak at about $6$ \si{\electronvolt} below the Fermi level consists mainly of O2p photoelectrons plus the underlying Si valence band \cite{keister_band_1999}. Our attention focuses on the O2s photoelectron peak at about $25$ \si{\electronvolt} binding energy. After subtraction of a Shirley background \cite{shirley_high-resolution_1972}, the integral O2s peak area is defined as the photoelectron yield $Y_{exp} \left( \theta \right)$ of the oxygen atoms in the SiO$_2$ layer at the sample surface, measured at a given angle $\theta$. Each $Y_{exp} \left( \theta \right)$ needs to undergo several normalization steps described in detail in Appendix \ref{electron-yield-normalization}. Importantly, the spectrum shown in Fig. \ref{Figure-2}(b) was measured in 20 minutes. To obtain a spectrum of similar statistics at any other XFEL, delivering hard x-ray single pulses at a maximum repetition rate of $120$ \si{\Hz} (for example, at the present LCLS), 9 hours of acquisition time would be needed. This makes time-resolved photoelectron spectroscopy measurements in the (sub)-ps time scale, without space charge effects, and with good statistics feasible only at high repetition rate XFELs, such as FLASH \cite{ackermann_operation_2007}, the European XFEL \cite{Altarelli2006} and LCLS-II \cite{stohr_linac_2011}.

\subsection{Photoelectron yield profiles} \label{Photoelectron yield profiles}

\begin{figure}[t]
\includegraphics[scale = 0.5]{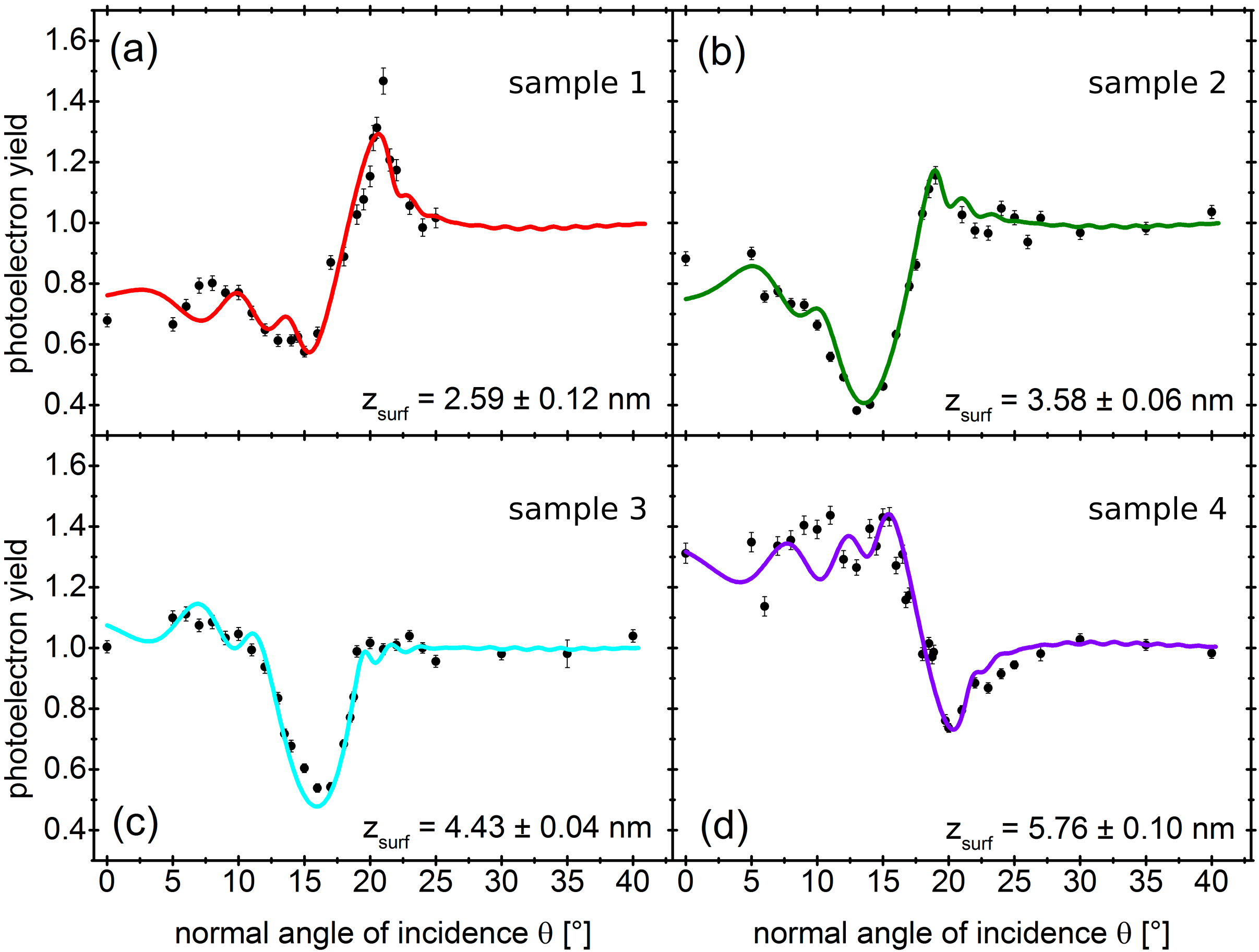}
\caption{\label{Figure-3} O2s photoelectron yield data $Y_{exp}\left(\theta\right)$ (black dots) and fit model $Y_{model}\left(\theta\right)$ (solid lines) for ML samples 1 (red, a), 2 (green, b), 3 (cyan, c), 4 (violet, d) with different $d_\mathrm{Si}^{top}$. The resulting fit parameter $z_{surf}$ is reported in each panel.}
\end{figure}

The structural information of XSW measurements is contained in the photoelectron yield profile, i.e. the sequence of $Y_{exp} \left( \theta \right)$ measured as the incidence angle $\theta$ is scanned through the Bragg condition. The normalized photoelectron yield profiles (Appendix \ref{electron-yield-normalization}) corresponding to ML samples with four nominally different top Si layers $d^{top}_\mathrm{Si}$ ($2.0$ \si{\nm}, $2.8$ \si{\nm}, $3.6$ \si{\nm} and $4.3$ \si{\nm}) are displayed in Fig. \ref{Figure-3}. The variations in yield follow from the XSW intensity variations at the top SiO$_2$ surface of each sample. Notably, the photoelectron yield profiles in Fig. \ref{Figure-3} are very different from each other and are strongly correlated with the thickness of the top Si layer. This indicates significantly different positions of the corresponding emitting oxygen atoms with respect to the standing wave modulation.

Importantly, the XSW effect can be exploited, by simply rotating the sample and thereby tuning the angle of incidence, to change the x-ray intensity within and above the sample, in this case by a factor of 3 (Fig. \ref{Figure-3}), without changing any of the beamline parameters. This feature can be very useful for a fast and reproducible fine tuning of the XFEL intensity at specific sample positions. 

\subsection{Photoelectron yield fit model} \label{Photoelectron yield fit model}


In order to extract the exact position of the O atoms contributing to the O2s photoelectron spectra we fitted the yield profiles with the model introduced below. First, we need to determine the relation between the intensity of the XSW and the measured photoelectron yield. In general, the photoelectron yield $Y \left( \theta \right)$ of an atom at a given position $z$ is not simply proportional to the XSW intensity $I_{SW}\left( z \right)$ as expressed in Eq. \eqref{eq_1}. In fact, for angularly resolved photoelectron spectroscopy in $\pi$-polarization the photoelectric cross-section in presence of an XSW depends on the experimental geometry. Particularly important are the direction and polarization of the x-ray waves and the direction of the emitted photoelectrons \cite{Vartanyants13TheX-RayStandingWaveTechnique:PrinciplesandApplications}. For our case of $\pi$-polarization the incident and Bragg-diffracted polarization vectors $\mathbf{e}_0$ and $\mathbf{e}_\mathrm{H}$ lie within the scattering plane, defined by the incident and Bragg-diffracted propagation vectors $\mathbf{k}_0$ and $\mathbf{k}_\mathrm{H}$, as shown in Fig. \ref{Figure-2}(c). Since soft x-rays are employed, in the calculation of the photoelectric cross-section  higher order multipole terms can be neglected \cite{vartanyants_non-dipole_2005}. Therefore, in the dipole approximation, for an initial s-state and in $\pi$-polarization geometry the angularly resolved photoelectron yield can be expressed as 
\begin{equation} \label{eq_2}
Y \left( \theta \right) = 1 + g^2 R\left(\theta\right) + 2 g \sqrt{R\left(\theta\right)} F_c \cos \left( \phi\left(\theta\right) - 2\pi P_c \right),
\end{equation}
where $g = \cos \theta_\mathrm{H} / \cos \theta_0 $ is the geometrical factor, with $\theta_0$ and $\theta_\mathrm{H}$ the angles between the polarization directions $\mathbf{e}_0$ and $\mathbf{e}_\mathrm{H}$ and the direction of the emitted electrons $\mathbf{n}_p$ (see insets in Fig. \ref{Figure-2}(c)). The coherent position is $P_c = \langle z\rangle/d_{SW}$, with $\langle z\rangle$ the average position of the emitting atoms contributing to the photoelectron yield, and the coherent fraction is $F_c$ indicating the distribution width of the emitters around their average position $\langle z\rangle$. Eq. \eqref{eq_2} is accurate if the distribution of atoms contributing to $Y$ does not extend for more than one layer and it is located at the sample surface, because in that case the damping of photoelectrons due to the inelastic mean free path can be neglected.


In our case, the measured photoelectron yield results from oxygen atoms in the top SiO$_2$ layer extending for $d_{\mathrm{SiO}_2} = 1.2$ \si{\nm} (Appendix \ref{SiO2 thickness}) below the surface $z_{surf}$, with $z = 0$ defined at the top of the Mo layer (see Figs. \ref{Figure-1}(c) and \ref{Figure-6}). Because of the inelastic mean free path, photoelectrons emitted from atoms below the surface at $z < z_{surf}$ and at a given angle $\alpha$ with the surface will contribute less to $Y\left( \theta \right)$ by a factor $e^{ -\left( z_{surf} - z \right) / \left( \lambda_{\mathrm{I}} \cdot \sin \alpha \right) }$ \cite{doring_standing-wave_2009}, where $\lambda_\mathrm{I}$ is the electron inelastic mean free path (Appendix \ref{electron-yield-normalization}) and $\alpha$ is the angle between the electron detection direction $\mathbf{n}_p$ and the sample surface (Fig. \ref{Figure-2}(c)). As a result, the fit model for $Y_{exp} \left( \theta \right)$ can be expressed as:
\begin{equation} \label{eq_3}
Y_{model} \left( \theta \right) = 1 + g_1 R\left(\theta\right) \\
+ 2 g_2 \sqrt{R\left(\theta\right)}\,\, \dfrac{\int\limits_{z_{surf} - d_\mathrm{SiO_2}}^{z_{surf}} e^{ -\left( z_{surf} - z \right) / \left( \lambda_{\mathrm{I}} \cdot \sin \alpha \right) } \cos\left( \phi\left(\theta\right) - 2\pi \frac{z}{d_{SW}}  \right) dz}{\int\limits_{z_{surf} - d_\mathrm{SiO_2}}^{z_{surf}} e^{ -\left( z_{surf} - z \right) / \left( \lambda_{\mathrm{I}} \cdot \sin \alpha \right)} dz}.
\end{equation}
Eq. \eqref{eq_3} represents the sum of photoelectron yield contributions between $z_{surf} - d_{\mathrm{SiO}_2}$ and $z_{surf}$ weighted by the inelastic mean free path factor. Since the geometrical factor $g$ in Eq. \eqref{eq_2} depends on the direction of the emitted photoelectrons, the acceptance angle $\pm 6 \degree$ of the time-of-flight spectrometer needs to be taken into account. Therefore, the geometrical factors $g_1$ and $g_2$ in Eq. \eqref{eq_3} are defined as follows:
\begin{equation}
g_1 \left( \theta \right) = \frac{1}{12\cos^2\theta_0} \int\limits_{-6}^{+6} \cos^2\left(\theta_\mathrm{H}+\theta_{\mathbf{n}_p}\right) d\theta_{\mathbf{n}_p},
\end{equation}
\begin{equation}
g_2 \left( \theta \right) = \frac{1}{12\cos\theta_0} \int\limits_{-6}^{+6} \cos\left(\theta_\mathrm{H}+\theta_{\mathbf{n}_p}\right) d\theta_{\mathbf{n}_p},
\end{equation}
where $\theta_{\mathbf{n}_p}$ indicates the emission angle relative to $\theta_\mathrm{H}$. The geometrical factors $g_1 \left( \theta \right)$ and $g_2 \left( \theta \right)$ depend on the normal angle of incidence $\theta$ via the angle $\theta_\mathrm{H} = 35\degree + 2\theta$ (Fig. \ref{Figure-2}(c)).


\subsection{Reflectivity data} \label{Reflectivity data}

\begin{figure}[t]
\includegraphics[scale = 0.8]{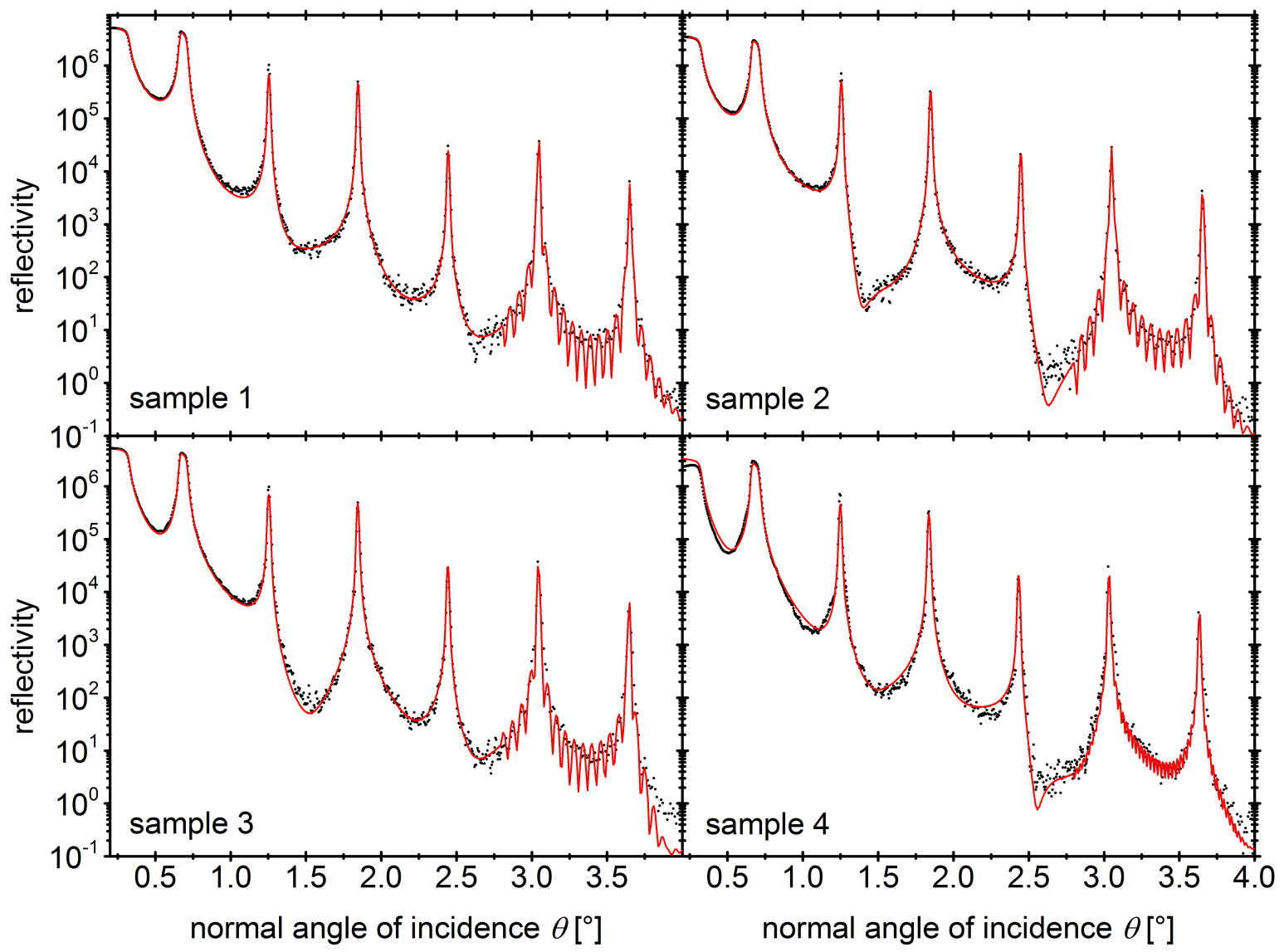}
\caption{\label{Figure-4} Experimental grazing incidence x-ray reflectivity data (black dots) and fit curves (red line) of samples 1, 2, 3, and 4, resulting from the combined fit of GIXR and EUVR data.}
\end{figure}

\begin{figure}[t]
\includegraphics[scale = 1.2]{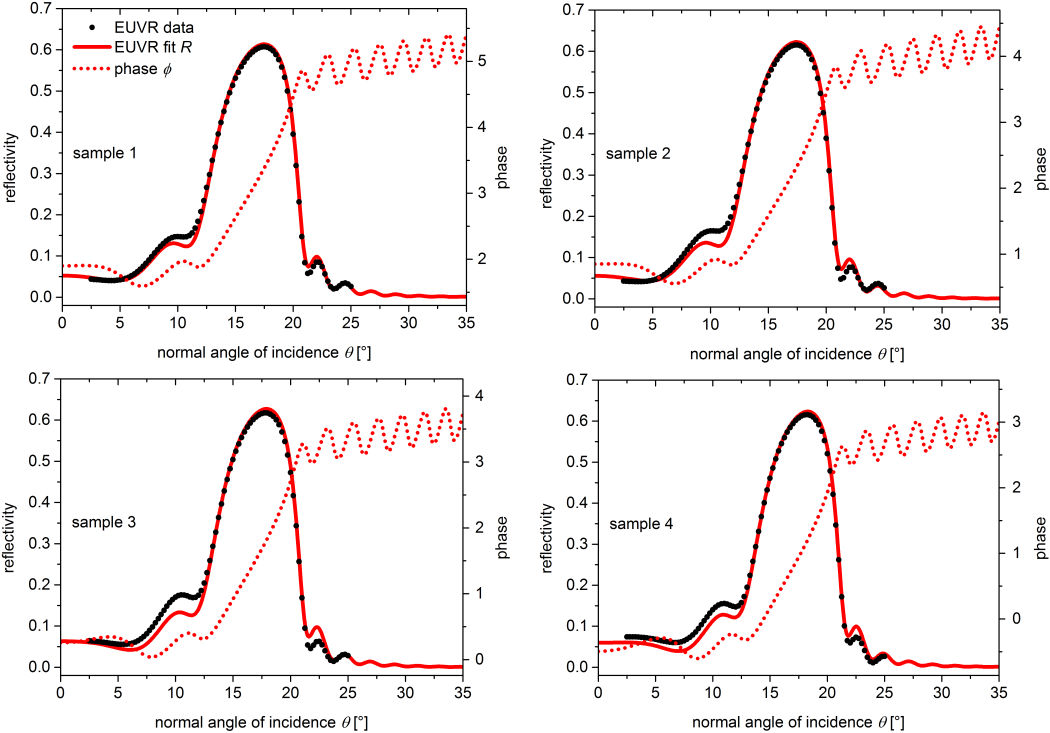}
\caption{\label{Figure-5} Experimental EUV reflectivity data (black dots), fit curves (red line) of samples 1, 2, 3, and 4, resulting from the combined fit of GIXR and EUVR data, and the corresponding phase $\phi\left(\theta\right)$ (red dashed line) calculated at the top of the SiO$_2$ layer in the ML samples. Note that the different scales of the the phase term result from the different positions of the top SiO$_2$ layer with respect to the periodic ML structure in samples 1 to 4 (Fig. \ref{Figure-6}).}
\end{figure}

To apply Eq. \eqref{eq_3} it is necessary to know the reflectivity $R\left(\theta\right)$, the phase $\phi\left(\theta\right)$ of the complex electric field amplitude ratio $E_\mathrm{H}/E_0$, and the period of the standing wave $d_{SW}$. These parameters could be easily calculated if the exact structure of our multilayer samples was known. To determine these parameters, grazing incidence x-ray reflectivity (GIXR, $\lambda = 0.154$ \si{\nm}) and extreme ultraviolet reflectivity (EUVR, $\lambda = 13.5$ \si{\nm}) measurements were performed using respectively a laboratory Cu K$_\alpha$ source (PANalytical Empyrean) at the University of Twente and the Metrology Light Source synchrotron radiation at the Physikalisch-Technische Bundesanstalt (PTB) in Berlin on the same samples probed with XSW at FLASH. GIXR and EUVR data are reported in Figs. \ref{Figure-4} and \ref{Figure-5} together with the corresponding fitting curves (obtained as described below) and phase calculations.

First, using the assumption independent approach \cite{zameshin_reconstruction_2016} GIXR measurements from sample 2 with $d_\mathrm{Si}^{top} = 2.8$ \si{\nm} were analyzed. The Mo/Si bilayer in the repetitive part of the multilayer and the top Mo/Si bilayer were divided in 30 sub-layers of Mo$_{1-x}$Si$_{x}$, where $x$ was a fitting parameter. The best fit model was parameterized introducing Mo and Si layers, Mo$_{1-x}$Si$_{x}$ interlayers, and sinusoidal transition layers between them for the simultaneous fit of GIXR and EUVR data as described in Ref.\cite{yakunin_combined_2014}. Second, the data of samples 1, 3, and 4 were fitted using the same 49 Mo/Si bilayers derived from sample 2. The periodic ML structure was identical for all samples as a result of the coating procedure. The only fitting parameters were: the thickness of the top Si layer $d_\mathrm{Si}^{top}$, the thickness of the transition from SiO$_2$ to vacuum, the total period thickness $d_\mathrm{ML}$. The simultaneous best fit of GIXR and EUVR data \cite{yakunin_combined_2014} provided a structural model for each of the four samples with different $d_\mathrm{Si}^{top}$. The real part of the refractive index $\delta$ for the top Mo/Si bilayers and the SiO$_2$ above is displayed in Fig. \ref{Figure-6}. The resulting structural parameters of the identical ML periodic structure are $d_\mathrm{Mo} = 3.3$ \si{\nm} and $d_\mathrm{Si} = 4.0$ \si{\nm}, leading to a standing wave periodicity of $d_{SW} = 7.3$ \si{\nm}.


Fits of EUVR data of each sample are reported in Fig. \ref{Figure-5} together with the calculation of the corresponding phase $\phi\left(\theta\right)$. The large and broad reflectivity peak with maximum of $61.4 \%$ at $\theta_{max} = 17.5\degree$ results from the Mo/Si ML, while the smaller side peaks, so-called Kiessig fringes, result from the interference of x-ray waves reflected at the vacuum-surface interface and ML-substrate interface. As the angle $\theta$ crosses the Bragg condition the phase $\phi\left(\theta\right)$ experiences a total variation of $\pi$, corresponding to a total shift of the XSW by $d_{SW}/2$, hence leading to the photoelectron yield modulations reported in Fig. \ref{Figure-3}. The phase term $\phi\left(\theta\right)$ was calculated at the top of the SiO$_2$ layer, therefore at different positions with respect to the periodic ML structure (Fig. \ref{Figure-6}). This results into rigid phase shifts going from sample 1 to 4 as it is evident from the corresponding scales in Fig. \ref{Figure-5}.

\subsection{Discussion} \label{Discussion}

The good quality of EUVR and GIXR curve fits enable us to employ the corresponding $R\left(\theta\right)$ and $\phi\left(\theta\right)$ functions to fit the experimental yield data $Y_{exp}\left(\theta\right)$ by means of the model in Eq. \eqref{eq_3}. The results summarized in Fig. \ref{Figure-3} show that $Y_{model}\left(\theta\right)$ describes very well our measured data. Two fit parameters were employed: the position of SiO$_2$ surface $z_{surf}$ and the angular offset to account for the slightly different angular scales of reflectivity and photoelectron yield measurements. The surface of the SiO$_2$ layer $z_{surf}$ in samples 1 to 4 was found to be respectively at $2.59 \pm 0.12$ \si{\nm}, $3.58 \pm 0.06$ \si{\nm}, $4.43 \pm 0.04$ \si{\nm}, and $5.76 \pm 0.10$ \si{\nm} above the top Mo layer, while the angular offset was of about $1.5 \degree$. 

The increase of $z_{surf}$ going from sample 1 to 4 follows directly from the larger $d_\mathrm{Si}^{top}$ leading to an increasing distance of the sample surface from the periodic ML structure as illustrated in Fig. \ref{Figure-6}. In this way we demonstrate the structural sensitivity of the XSW technique using XFEL pulses. In particular, the small error bars (Appendix \ref{Error_analysis}) of $z_{surf}$ ($<0.15$ \si{\nm}) indicate the high spatial accuracy of the measured SiO$_2$ surface positions.


\section{Conclusion} \label{Conclusion}

In this letter we have demonstrated the structural sensitivity and accuracy of the XSW technique at an XFEL. In combination with the high chemical specificity and surface sensitivity of photoelectron spectroscopy and together with the femtosecond duration of XFEL pulses, these experiments open up the possibility of obtaining direct ultrafast structural information of reactants involved in chemical reactions at surfaces. Time-resolved structural data will enormously contribute to the fundamental understanding of more complex processes in heterogeneous catalysis both on single metal crystals \cite{nilsson_catalysis_2017} and more exotic layered crystals as perovskites \cite{hwang_perovskites_2017}, leading eventually to more efficient catalysts. In addition, time-resolved XSW may reveal the structural dynamics at the basis of light-induced superconductivity \cite{fausti_light-induced_2011, hu_optically_2014, mankowsky_nonlinear_2014} by providing element and site specific atomic positions \cite{kovalchuk_observation_1997, kazimirov_x-ray_2000, thiess_resolving_2015} as a function of the delay from the light pump pulse. This could pave the way to solve the longstanding puzzle of high critical temperature superconductors and indicate the appropriate crystal structure to enhance superconductivity. 


\begin{acknowledgements}

We acknowledge the support of FLASH scientific and technical staff for making the experiment possible. We are grateful to T. Kroesen for help in the XPS laboratory measurements. This work is supported by the Deutsche Forschungsgemeinschaft within the excellence cluster “Center for Ultrafast Imaging (CUI).” GM, IAV and WW acknowledge partial funding from Helmholtz-Russia grant on "Big data". IAM and IM acknowledge the support of the Industrial Focus Group XUV Optics, MESA+ Institute for Nanotechnology, University of Twente, notably the industrial partners ASML, Carl Zeiss SMT GmbH, PANalytical, as well as the Province of Overijssel and the NWO. We thank S. N. Yakunin (Kurchatov Institute, Moscow, Russia) for help with simulations and useful discussions. 

\end{acknowledgements}

\appendix

\section{Photoelectron yield normalization} \label{electron-yield-normalization}

Each $Y_{exp} \left( \theta \right)$ needs to undergo the following normalization steps before fitting the photoelectron yield profile using the fit model of Eq. \eqref{eq_3} described in section \ref{Photoelectron yield fit model}.

\paragraph{Normalization by the XFEL intensity and acquisition time.} The intensity of a SASE (Self-Amplified Spontaneous Emission) XFEL varies from pulse to pulse with variations up to approximately $20\%$, therefore it is necessary to normalize each electron yield data point $Y_{exp}\left(\theta\right)$ by the corresponding XFEL intensity. As a reference for the XFEL intensity we consider the ion signal of a gas monitor detector \cite{tiedtke_soft_2009} located directly after the undulators of FLASH (Fig. \ref{Figure-2}(a)). The normalization factor was calculated as the sum of the ion signal over the entire acquisition run. In this way, not only we normalize by the incoming XFEL intensity but also by the acquisition time.

\paragraph{Normalization by the filter transmission.} \label{subsecton-Filters} After the gas monitor detector and before the monochromator of PG2 beamline at FLASH there is a gas absorber and several solid filters that can be used to reduce the XFEL intensity. The pressure of Xe in the gas absorber was always kept constant to \SI{2.7e-2}{\milli\bar}, hence normalization by the corresponding attenuation factor is not necessary. In contrast, some of the solid filters were used and changed during the acquisition of electron yield data of the same yield profile as reported in Table \ref{table-01-Filter}. Therefore, in order to have consistent data within the same yield profile each electron yield $Y_{exp}\left(\theta\right)$ needs to be normalized by the corresponding filter transmission. We kept a Si$_3$N$_4$ filter $500$ \si{\nm} thick throughout all the measurements, while we alternated two ZrB$_2$ filters with thickness $431$ \si{\nm} and $200$ \si{\nm}. The last two filters were used either both in series or only one of the two as indicated in Table \ref{table-01-Filter}, where also the corresponding photon energy is reported.

\begin{table}[h]
\begin{ruledtabular}
\begin{tabular}{cccc}
Settings & Sample (angular range [$\degree$]) & Photon energy [\si{\eV}] & Filters used\\
\colrule
1 & 1 (0-17), 4 (0-25) & 93.0 & 1+2\\
2 & 1 (18-25) & 93.0 & 1+2+3\\
3 & 2 (0-18), 3 (0-25) & 91.7 & 1+3\\
4 & 2 (18.5-40), 3 (30-40), 4 (27-40) & 91.7 & 1+2\\
\end{tabular}
\end{ruledtabular}
\caption{\label{table-01-Filter} Sample number, angular range, photon energy and filter configuration used for XSW measurements are reported. Filters 1, 2, and 3 refer to Si$_3$N$_4$ $500$ \si{\nm}, ZrB$_2$ $431$ \si{\nm}, and ZrB$_2$ $200$ \si{\nm} respectively.}
\end{table}

\paragraph{Normalization by the inelastic mean free path factor.} \label{IMFP} Since the electron yied $Y_{exp} \left( \theta \right)$ is measured at different angles of incidence $\theta$, the number of photoelectrons that can leave the surface and reach the detector varies depending on the effective number of atomic layers crossed by the photoelectrons. The factor denoting the damping of the emitted O2s photoelectrons due to the corresponding inelastic mean free path $\lambda_I$ is $I\left(z,\theta\right) = e^{-\left( z_{surf} - z \right)/\left( \lambda_I \cdot \sin \alpha \right)}$, where $\alpha \left( \theta \right)$ is the angle between the surface and the direction $\mathbf{n}_p$ of photoemitted electrons towards the detector (Fig. \ref{Figure-2}(c)). Since the angle between $\mathbf{k}_0$ and $\mathbf{n}_p$ is $55 \degree$, it follows that $\alpha = 35 \degree + \theta$. Following the notation used in the article, the coordinate $z$ indicates positions perpendicular to the sample layers with $z=0$ at the top of the last Mo layer, $z>0$ above it (Figs. \ref{Figure-1}(c) and \ref{Figure-6}), and $z_{surf}$ is defined as the position of the SiO$_2$ surface. The normalization factor accounting for $\lambda_I$ is given by
\begin{equation}
N_{\lambda}\left(\theta\right) = \int\limits_{z_{surf} - d_\mathrm{SiO_2}}^{z_{surf}} e^{-\left( z_{surf} - z \right)/\left[ \lambda_I \cdot \sin \left( 35 \degree + \theta \right) \right]} dz,
\end{equation}
where $d_\mathrm{SiO_2} = 1.2$ \si{\nm} is the average thickness of the SiO$_2$ layer, as measured by x-ray photoelectron spectroscopy (Appendix \ref{SiO2 thickness}). The normalization factor $N_{\lambda}\left(\theta\right)$ was calculated for each of the angle of incidence $\theta$ at which experiments were performed. The inelastic mean free path $\lambda_\mathrm{I} = 4.4$ \si{\angstrom} results from the interpolation of tabulated values obtained from the TPP-2 formula \cite{tanuma_calculations_1991-1} and it was calculated for O2s photoelectrons with kinetic energy $62$ \si{\electronvolt} going through SiO$_2$.

\paragraph{Normalization by the XFEL footprint.} We need to consider that for $\theta > 0 \degree$ the footprint of the XFEL at the sample will increase in the horizontal direction by a factor $1 / \cos \theta $, thus a larger number of photoelectrons will be detected. To account for this geometrical factor, $Y_{exp} \left( \theta \right)$ is normalized by $1 / \cos \theta$. 

\paragraph{Normalization by the photoelectron yield off Bragg.} Finally, each yield profile is normalized by the photoelectron yield measured away from Bragg condition. From simulations of yield profiles it results that independently from the position of the emitting atoms (for any $P_c$) for $\theta > 27\degree$ the yield $Y_{exp} \left( \theta \right)$ (normalized to the intensity of the incoming x-ray electric field $|E_0|^2$) differs from 1 by less than $1\%$. Therefore, the yield data of each sample are normalized by the corresponding average of $Y_{exp} \left( \theta \right)$ for $\theta > 27\degree$. In the case of sample 1 with nominal $d_{\mathrm{Si}}^{top} = 2.0$ \si{\nm}, since experimental data are available only up to $\theta = 25\degree$, this normalization factor is an additional fit parameter (Appendix \ref{Error_analysis}).

\section{Error analysis.} \label{Error_analysis} The error bars of the SiO$_2$ surface positions $z_{surf}$ reported in the article are calculated as in Ref.\cite{mercurio_adsorption_2013}. First, the statistical error of each measured electron yield is calculated as the standard deviation of $400$ synthetic spectra, generated by Monte Carlo simulations from each measured spectrum, assuming that the noise in the photoemission spectrum follows the Poisson distribution. Second, the fit of the electron yield profile by means of the Levenberg-Marquardt method yields a covariance matrix. The square root of the covariance matrix diagonal values are the error bars of the fit parameters $z_{surf}$, angular offset between reflectivity and photoelectron yield data, and photoelectron yield off Bragg (only for sample 1).

\section{Thickness of the top $\mathrm{SiO_2}$ layer} \label{SiO2 thickness}

\begin{figure}[t]
\includegraphics[scale = 0.9]{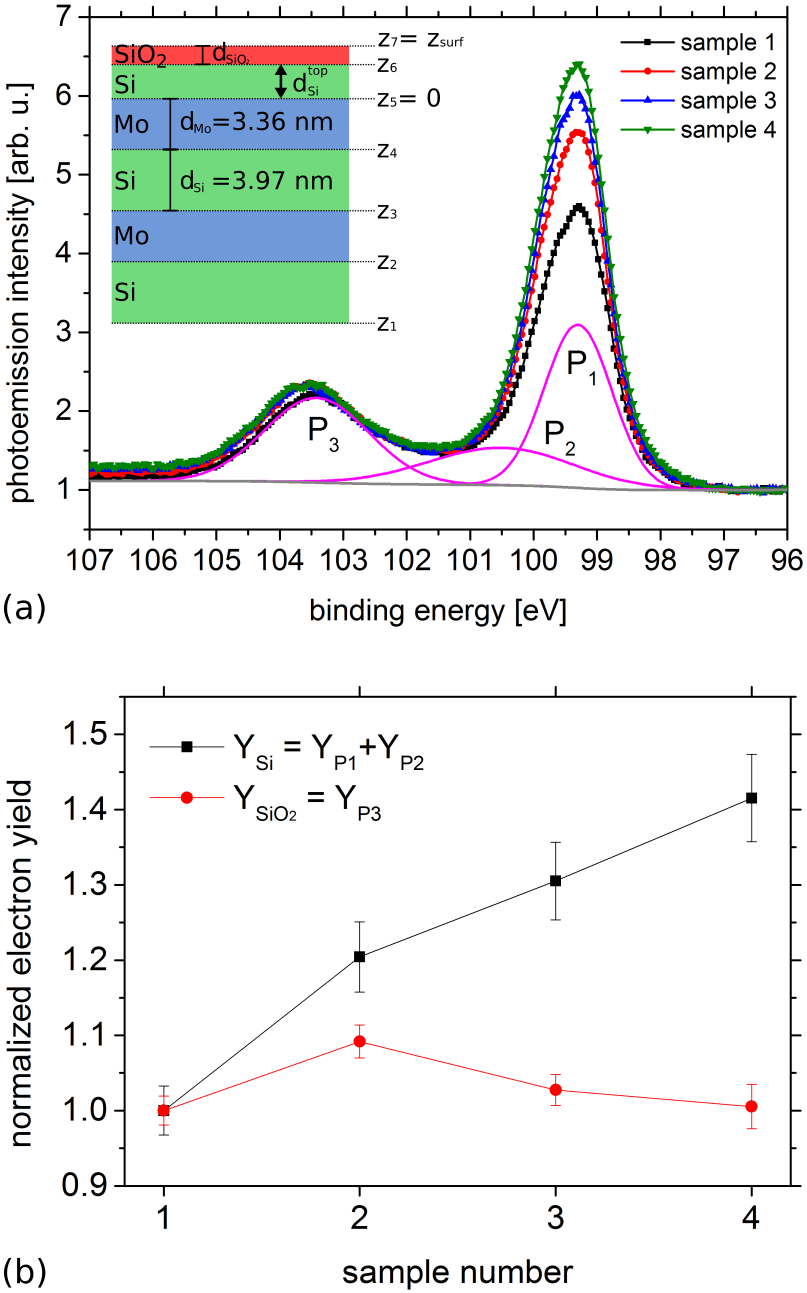}
\caption{\label{Figure-7} (a) Si2p PE spectra of Mo/Si ML samples 1, 2, 3, and 4 with different nominal thickness of the top Si layer $d_\mathrm{Si}^{top} = 2.0$ \si{\nm}, $2.8$ \si{\nm}, $3.6$ \si{\nm}, and $4.3$ \si{\nm}. Spectra are normalized to the background PE intensity at $93.7$ \si{\electronvolt}. Fitting gaussian components P$_1$, P$_2$ and P$_3$ of sample 1 PE spectrum are marked in magenta. The Shirley background is marked in gray. Inset: Sketch of 2 ML periods together with the top Si layer and the native SiO$_2$ layer. Thicknesses $d_\mathrm{Mo} = 3.36$ \si{\nm}, $d_\mathrm{Si} = 3.97$ \si{\nm} and $d_\mathrm{Si}^{top} = 1.82$ \si{\nm}, $2.56$ \si{\nm}, $3.57$ \si{\nm}, $4.23$ \si{\nm} for samples 1, 2, 3, 4 result from the combined analysis of GIXR and EUVR data (see section \ref{Reflectivity data}). The coordinates $z_i$, with $i=1,...,7$, indicate the positions of the interface between two layers, with $z_5 = 0$ defined as the origin of the coordinate $z$, and $z_7 = z_{surf}$ defined as the surface position of the top SiO$_2$ layer. (b) Normalized photoelectron yield of bulk-like Si ($Y_\mathrm{Si}$, black) and of SiO$_2$ ($Y_\mathrm{SiO_2}$, red).}
\end{figure}

The ML samples investigated by GIXR, EUVR and XSW were also measured by x-ray photoemission spectroscopy (XPS) with the aim to determine the thickness of the top native SiO$_2$ layer $d_\mathrm{SiO_2}$. Photoemission (PE) spectra were measured at the University of Hamburg by means of a hemispherical analyser Scienta SES-2002 and using the Mg K$\alpha$ radiation at $1253.6$ \si{\electronvolt} with $\mathrm{FWHM} = 1.5$ \si{\electronvolt}. Si2p PE spectra are displayed in Fig. \ref{Figure-7}(a), after normalization by the PE intensity of the background at $93.7$ \si{\electronvolt}. Each spectrum is fitted with three gaussian functions: P$_1$, P$_2$ and P$_3$. Component P$_1$ has binding energy (BE) $99.3$ \si{\electronvolt} and $\mathrm{FWHM} = 1.2$ \si{\electronvolt}, hence it is assigned to Si2p photoelectrons of bulk-like Si. On the other hand, component P$_3$ has $\mathrm{BE} = 103.5$ \si{\electronvolt} and $\mathrm{FWHM} = 2.0$ \si{\electronvolt}, therefore it is assigned to Si$^{+4}$ atoms in the native SiO$_2$ layer. This interpretation is supported by the average binding energy shift (P$_3$ - P$_1$) of $4.2$ \si{\electronvolt}, in agreement with previous results \cite{GrunthanerMSR86, himpsel_microscopic_1988, iwata_electron_1996}. The component P$_2$ relates to Si atoms with an oxidation state larger than 0 (Si in bulk) and smaller than +4 (Si in SiO$_2$). These Si atoms form a transition layer SiO$_x$ between Si and SiO$_2$ layers. Since the BE shift of P$_2$ with respect to P$_1$ is smaller than $1.2$ \si{\electronvolt} we conclude that SiO$_x$ mainly consists of Si atoms with +1 or smaller oxidation state \cite{iwata_electron_1996}, therefore close to bulk-like Si.

\begin{figure}[t]
\includegraphics[scale = 0.9]{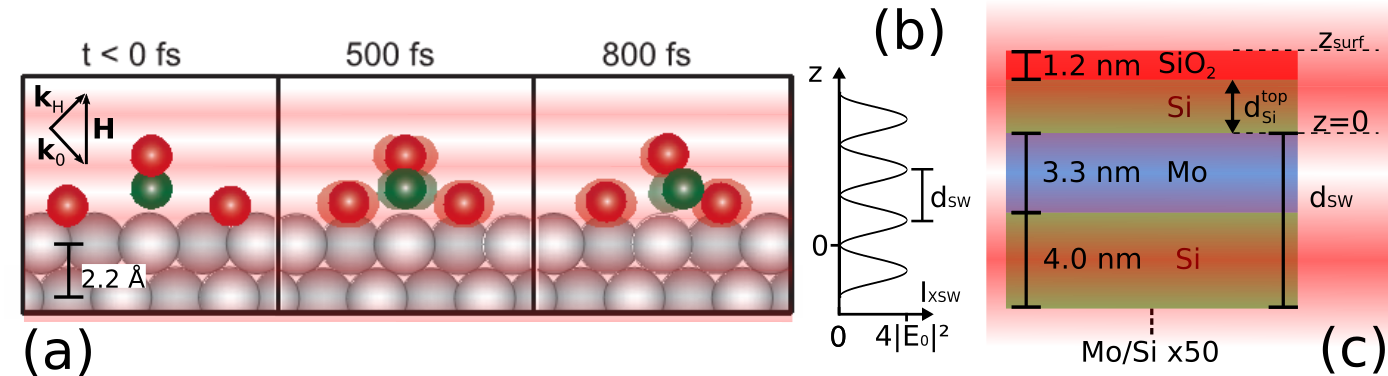}
\caption{\label{Figure-1} (a) Schematic sequence of CO oxydation reaction on a Ru surface (reproduced from Ref.\cite{ostrom_probing_2015}). Atoms coloured by gray, red, green correspond to Ru, O, C atoms. Inset: Bragg diffraction vector $\mathbf{H}$, incident $\mathbf{k}_0$ and Bragg-diffracted $\mathbf{k}_\mathrm{H}$ wave vectors. (b) XSW intensity $I_\mathrm{XSW}$ with maximum $4\left|E_0 \right|^2$, where $E_0$ is the electric field amplitude of the incoming x-ray wave, and $d_{SW}$ is the period of the standing wave. (c) Sketch of Mo, Si and SiO$_2$ top layers of the ML samples including the thicknesses $d_\mathrm{Mo} = 3.3$ \si{\nm} and $d_\mathrm{Si} = 4.0$ \si{\nm}, the period of the standing wave $d_{SW} = 7.3$ \si{\nm}, the variable thickness of the top Si layer $d_\mathrm{Si}^{top}$, and the position of surface O atoms ${z}_{surf}$ with respect the top of the Mo layer at $z = 0$.}
\end{figure}

\begin{figure}[t]
\includegraphics[scale = 0.9]{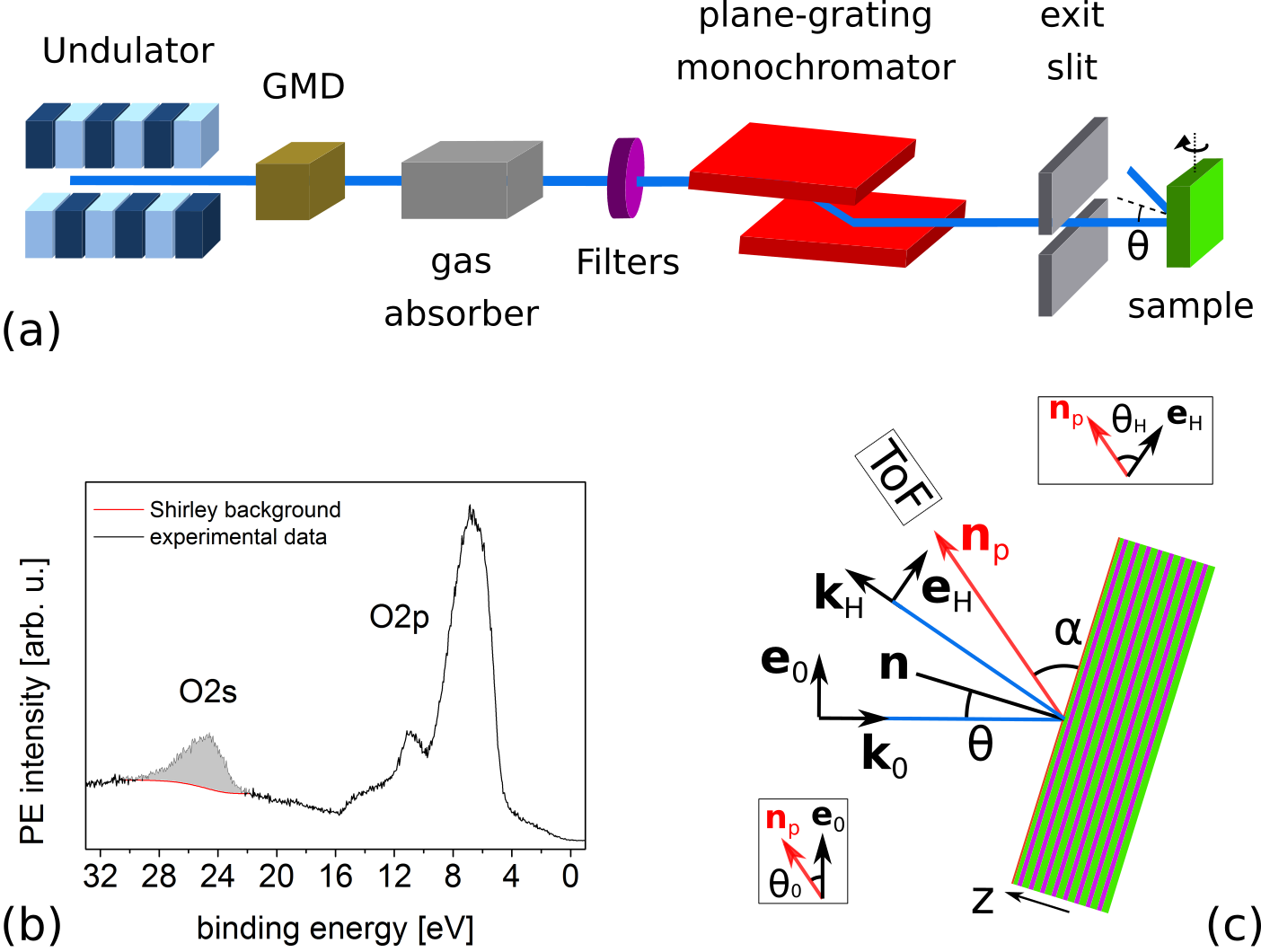}
\caption{\label{Figure-2} (a) Simplified scheme of FLASH PG2 beamline \cite{martins_monochromator_2006} including undulator, gas monitor detector (GMD), gas absorber, solid filters, plane-grating monochromator (with 200 lines/mm, operated in first order), exit slit (set to $100$ \si{\um}) and rotatable sample. (b) Photoelectron spectrum of O2s and O2p lines measured on a Mo/Si ML sample at $\theta = 0\degree$. The oxygen electron yield is marked by the gray area. (c) Top view of a ML sample and Time of Flight (ToF) spectrometer with all relevant vectors and angles. The wave vectors and polarization vectors of the incident [Bragg-diffracted] x-ray wave are $\mathbf{k}_0$ and $\mathbf{e}_0$ [$\mathbf{k}_\mathrm{H}$ and $\mathbf{e}_\mathrm{H}$]. The normal angle of incidence $\theta$ is defined between $\mathbf{k}_0$ and the normal $\mathbf{n}$ to the ML surface. The vector $\mathbf{n}_p$ indicates the direction of the photoelectrons towards the ToF spectrometer. The angle $\alpha$ is defined between $\mathbf{n}_p$ and the sample surface. The angle between $\mathbf{e}_0$ [$\mathbf{e}_\mathrm{H}$] and $\mathbf{n}_p$ is $\theta_0$ [$\theta_\mathrm{H}$]. In our experimental geometry, $\theta_0 = 35\degree$ is constant, $\theta_\mathrm{H} = 35\degree + 2\theta$ and $\alpha = 35\degree + \theta$. The coordinate $z$ indicates positions perpendicular to the sample layers with $z>0$ above the ML and $z=0$ at the top Mo layer (Figs. \ref{Figure-1}(c) and \ref{Figure-6}).}
\end{figure}

\begin{figure}[t]
\includegraphics[scale = 0.5]{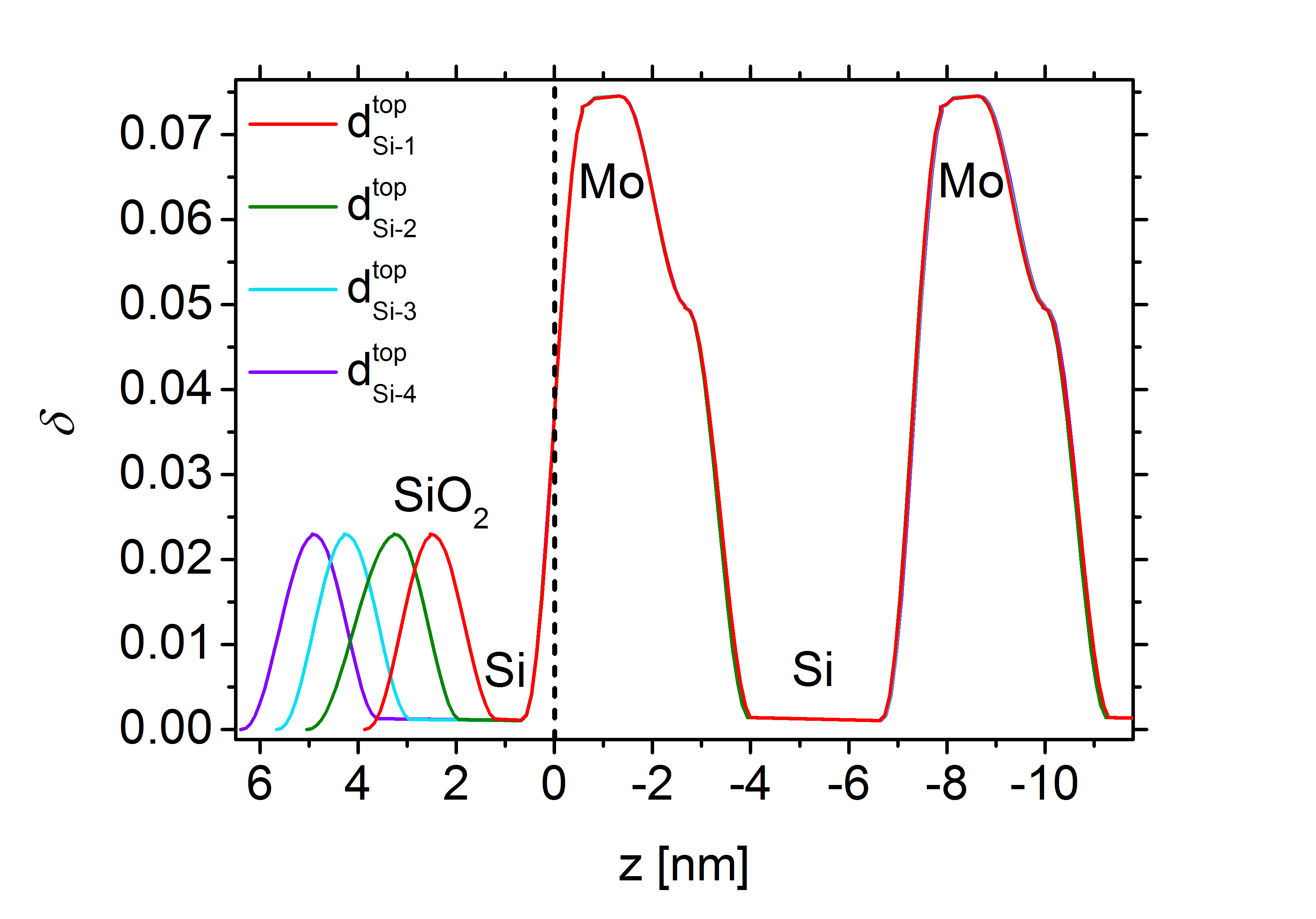}
\caption{\label{Figure-6} Real part of the refractive index $\delta$ of the top Mo/Si bilayers and the SiO$_2$ above, calculated for $\lambda = 13.5$ \si{\nm} and for different $d_\mathrm{Si}^{top}$ of samples 1 to 4. The obtained structural parameters are: $d_\mathrm{ML} = 7.33 \pm 0.07$ \si{\nm}, $d_\mathrm{Mo} = 3.36 \pm 0.03$ \si{\nm}, $d_\mathrm{Si} = 3.97 \pm 0.04$ \si{\nm}, $d_\mathrm{SiO_2} = 1.44 \pm 0.20$ \si{\nm} and $d_\mathrm{Si}^{top} = 1.82 \pm 0.20$ \si{\nm}, $2.56 \pm 0.30$ \si{\nm}, $3.57 \pm 0.40$ \si{\nm} and $4.23 \pm 0.40$ \si{\nm} for samples 1, 2, 3, and 4.}
\end{figure}

To determine the thickness of the SiO$_2$ layer, the photoelectron yield of bulk-like Si is defined as the sum of the yield of component P$_1$ and P$_2$, $Y_\mathrm{Si} = Y_\mathrm{P_1} + Y_\mathrm{P_2}$, while $Y_\mathrm{SiO_2} = Y_\mathrm{P_3}$ (Fig. \ref{Figure-7}(b)). The electron yield of Si, within the layer $y$ (with $y$ = Si, SiO$_2$), can be expressed as
\begin{equation}\label{eq-01}
Y_{y} = \sigma_\mathrm{Si} N_y Q_\mathrm{Si} \int_{z_i}^{z_{i+1}} e^{-\frac{z_{surf}-z}{\lambda_{\mathrm{Si},y}}} dz,
\end{equation}
where: $\sigma_\mathrm{Si}$ is photoionization cross section, $N_y = \rho_y N_A / M_y$ is the number density (in atoms \si{\per\cubic\centi\metre}) of the element/compound $y$, $\rho_y$ is the density (in \si{\gram\per\cubic\centi\metre}), $N_A$ is Avogadro's number, $M_y$ is the molar mass (in \si{\gram\per\mole}), $Q_{Si}$ is the transmission of the hemispherical analyser which depends on the kinetic energy of the photoelectrons, $\lambda_{Si,y}$ is the inelastic mean free path of Si photoelectrons in the $y$ layer, and $z$ is the coordinate perpendicular to the ML planes. The exponential term has at the denominator $\sin\alpha = 1$ since XPS measurements were carried out at normal emission ($\alpha = 90\degree$). Eq. \eqref{eq-01} refers to the yield of photoelectrons coming from a layer located between positions $z_i$ and $z_{i+1}$, with $i=1,...,7$, where $z_5 = 0$ is the origin of the $z$ coordinate (Figs. \ref{Figure-1}(c) and \ref{Figure-6}), and $z_7 = z_{surf}$ is the surface position of the top SiO$_2$ layer (see inset in Fig. \ref{Figure-7}(a)). Due to the very small BE energy difference of $4$ \si{\electronvolt} between Si2p photoelectrons from Si and SiO$_2$, the corresponding differences in photoionization cross section and analyzer transmission can be neglected. The number densities $N_\mathrm{Si} = 5.01$ \si{\per\cubic\centi\metre} and $N_\mathrm{SiO_2} = 2.66$ \si{\per\cubic\centi\metre} are calculated using molar masses $M_\mathrm{Si} = 28$ \si{\gram\per\mole}, $M_\mathrm{SiO_2} = 60$ \si{\gram\per\mole} and densities $\rho_\mathrm{Si} = 2.33$ \SI{e22}{\gram\per\cubic\centi\metre}, $\rho_\mathrm{SiO_2} = 2.65$ \SI{e22}{\gram\per\cubic\centi\metre}, respectively. The inelastic mean free path of Si electrons in a Si layer and in a SiO$_2$ layer is $\lambda_\mathrm{Si,Si} = 27.2$ \si{\angstrom} and $\lambda_\mathrm{Si,SiO_2} = 24.4$ \si{\angstrom}, resulting from the interpolation of values tabulated in Ref.\cite{tanuma_calculations_1991} and Ref.\cite{tanuma_calculations_1991-1} at the respective electron kinetic energies $1154$ \si{\electronvolt} and $1150$ \si{\electronvolt}.

The ratio of photoelectron yields $Y_\mathrm{SiO2}/Y_\mathrm{Si}$ can be expressed as:
\begin{equation}\label{eq-ratio}
\frac{Y_\mathrm{SiO_2}}{Y_\mathrm{Si}} = \frac{\sigma_\mathrm{Si} N_\mathrm{SiO_2} Q_\mathrm{Si} \int_{z_6}^{z_{surf}} e^{-\frac{z_{surf}-z}{\lambda_\mathrm{Si,SiO_2}}} dz}{\sigma_\mathrm{Si} N_\mathrm{Si} Q_\mathrm{Si} \int_{z_1}^{z_6} e^{-\frac{z_{surf}-z}{\lambda_\mathrm{Si,Si}}} dz},
\end{equation}
with the integrals in $z$ limited between the surface ($z_{surf}$) and the bottom of the third Si layer ($z_1$) (see inset in Fig. \ref{Figure-7}(a)) since deeper layers contribute to the electron yield by less than $0.1 \%$. Eq. \eqref{eq-ratio} can be recast as:
\begin{multline}\label{eq-explicit}
\frac{Y_{SiO_2}}{Y_{Si}} = \\ \frac{ N_\mathrm{SiO_2} \lambda_\mathrm{Si,SiO2} \left( e^{-\frac{z_{surf}-z_6}{\lambda_\mathrm{Si,SiO_2}}} - 1\right)}{N_\mathrm{Si} \lambda_\mathrm{Si,Si} \left[ \left( e^{-\frac{z_{surf}-z_1}{\lambda_\mathrm{Si,Si}}} - e^{-\frac{z_{surf}-z_2}{\lambda_{Si,Si}}} \right) + \left( e^{-\frac{z_{surf}-z_3}{\lambda_{Si,Si}}} - e^{-\frac{z_{surf}-z_4}{\lambda_{Si,Si}}} \right) + \left( e^{-\frac{z_{surf}-z_5}{\lambda_{Si,Si}}} - e^{-\frac{z_{surf}-z_6}{\lambda_{Si,Si}}} \right) \right]}.
\end{multline}
The thicknesses of Mo, Si and top Si layers are known from GIXR and EUVR analysis (section \ref{Reflectivity data}): $d_\mathrm{Mo} = 3.36$ \si{\nm}, $d_\mathrm{Si} = 3.97$ \si{\nm}, $d_\mathrm{Si}^{top} = 1.82$ \si{\nm}, $2.56$ \si{\nm}, $3.57$ \si{\nm}, and $4.23$ \si{\nm} for samples 1, 2, 3, and 4. The only unknown parameter is $d_\mathrm{SiO_2}$. By solving Eq. \eqref{eq-explicit} numerically using python, we obtain the following thicknesses of the native oxide layer $d_\mathrm{SiO_2}$ in samples 1 to 4: $1.18 \pm 0.04$ \si{\nm}, $1.22 \pm 0.04$ \si{\nm}, $1.21 \pm 0.04$ \si{\nm}, and $1.16 \pm 0.05$ \si{\nm}. The resulting average is $d_\mathrm{Si}^{top} = 1.19 \pm 0.04$ \si{\nm}.

\bibliographystyle{apsrev4-1}


%

\newpage

\end{document}